\documentclass[english,letterpaper,aps,prl,twocolumn,showpacs]{revtex4-1}
\usepackage{lmodern}

\usepackage[T1]{fontenc}
\usepackage[latin1]{inputenc}
\usepackage{graphicx}
\usepackage{amssymb}

\makeatletter

\usepackage{lmodern}
\usepackage{subfig}

\makeatletter

\usepackage{lmodern}

\makeatletter

\usepackage{lmodern}

\makeatletter

\usepackage{lmodern}

\makeatletter

\makeatletter

\usepackage{epsfig}

\usepackage{dcolumn}

\usepackage{bm}

\usepackage{bbm}

\usepackage{wasysym}

\newlength{\textwidthm}

\makeatother

\makeatother

\makeatother

\makeatother

\makeatother

\usepackage{babel}
\makeatother
\begin{document}

\title{RKKY interaction in gapped or doped graphene}

\author{E. Kogan}
\email{Eugene.Kogan@biu.ac.il}

\affiliation{Department of Physics, Bar-Ilan University, Ramat-Gan 52900,
Israel}
\date{\today}

\begin{abstract}
In our previous work (E. Kogan, Phys. Rev. B {\bf 84}, 115119 (2011)) we
 calculated RKKY interaction between two magnetic impurities in pristine graphene using the Green's functions (GF) in the coordinate -- imaginary time  representation. Now we show that the calculations of the GF in this representation can be simplified by using the  Feynman's trick, which allows to easily calculate  RKKY interaction in gapped graphene. We also present calculations of the RKKY interaction in gapped or doped graphene using
the coordinate -- imaginary frequency representation.
Both representations, corresponding to calculation of the bubble diagram in Euclidean space,
have an important advantage over those corresponding to calculation in Minkowskii space, which are  very briefly reviewed  in the Appendix to the present work. The former, in distinction to the  latter,  operate only with the convergent integrals from the start to the end of the calculation.
\end{abstract}

\pacs{75.30.Hx;75.10.Lp}

\maketitle

\section{Introduction}

RKKY interaction between two magnetic impurities in graphene was theoretically studied quite intensely during last several years
 \cite{vozmediano,dugaev,brey,saremi,cheianov,black,sherafati,uchoa,bunder,power,kogan,sherafat,szalowski,bunder2,kettemann,sherafati2,liwei,loss,gumbs}. (A terse but precise review
 of the issue one can find in the book by M. Katsnelson \cite{katsnelson}.)
One may ask, why  the problem, which is  in principle so simple (in the lowest order of perturbation theory, as
it was treated in all the papers referenced above, the problem is equivalent to calculation of a single bubble diagram), was the subject of so many publication, using different approaches?
One of the answers to this question is connected with the fact that a simply written integral is not necessarily a simply calculated  integral.
More specifically, in most of the approaches mentioned above  the integrals defining the RKKY interaction in graphene   turned out to be divergent,
and to obtain  finite values from these divergent integrals  one has
to implement the complicated (and to some extent arbitrary) cut-off procedure.  The cure from this disease  turned out to be calculation of this diagram not in Mikowskii, but in Euclidean space. Green's function (GF)  in  coordinate -- imaginary frequency representation were used in Ref. \cite{dugaev}. GF in   coordinate -- imaginary time representation were used in Refs. \cite{cheianov,kogan}.

In  the present work we generalize the approach of our previous publication \cite{kogan} to treat the case of gapped graphene.
Not to distract attention of the reader from the aspects of the physics we are going to concentrate upon,
we consider a toy model of graphene, with free electrons being described by the 2d Dirac Hamiltonian. The existence of two Dirac points in graphene leads to additional  angular dependent factor in the formula for the RKKY interaction. This angular dependence was thoroughly studied previously \cite{saremi,sherafati} and does not interfere with the physics we are discussing in this work.

The effective exchange  RKKY  interaction between the two magnetic impurities with the  spins ${\bf S}_1$ and ${\bf S}_2$,   sitting on top of carbon atoms with the relative radius--vector ${\bf R}$ is
\begin{eqnarray}
\label{abr5}
H_{RKKY}=-\frac{1}{4}J^2\chi(R){\bf S}_1{\bf\cdot  S}_2,
\end{eqnarray}
where $J$ is the contact exchange interaction between each of spins and the graphene electrons, and $\chi(R)$ is the free electrons charge  susceptibility, depending upon whether the carbon atoms belong to the same, or different sublattices.

\section{Imaginary time representation: gapped graphene}

The susceptibility expressed through the GF in  the imaginary time --coordinate representation  was written as  \cite{saremi,cheianov,kogan}
\begin{eqnarray}
\label{abr}
\chi(R)=-4\int_{0}^{\infty}{\cal G}({\bf R};\tau){\cal G}(-{\bf R};-\tau)d\tau.
\end{eqnarray}

Consider  electrons in 2d described by Dirac equation
\begin{eqnarray}
\label{dirac2}
\left[i\partial_{t}+i v{\bf \gamma\cdot \partial}-\gamma^0m\right]\psi=0,
\end{eqnarray}
where
${\bf \gamma\cdot \partial}=\gamma^1\partial_1+\gamma^2\partial_2$.
To find the GF in coordinate-time representation  it is convenient to use the trick introduced by Feynman  \cite{feynman},
that is to represent the GF  as
\begin{eqnarray}
\label{real1}
{\cal G}(R,t)=\left[i\partial_{t}+i v{\bf \gamma\cdot \partial}+\gamma^0m\right]I(R,\tau),
\end{eqnarray}
where
\begin{eqnarray}
\label{real2}
I(R,t)=\int \frac{d^2k}{(2\pi)^2}\frac{1}{2E_k}e^{-iE_k|t|+i{\bf k\cdot R}},
\end{eqnarray}
and $E_k=\sqrt{v^2k^2+m^2}$.

Going from real to  imaginary time, instead of Eqs. (\ref{real1}) and (\ref{real2}) we obtain respectively
\begin{eqnarray}
{\cal G}(R,\tau)=\left[\partial_{\tau}+i v{\bf \gamma\cdot \partial}+\gamma^0m\right]I(R,\tau),
\end{eqnarray}
and
\begin{eqnarray}
\label{imaginary2}
I(R,\tau)=\int \frac{d^2k}{(2\pi)^2}\frac{1}{2E_k}e^{-E_k|\tau|+i{\bf k\cdot R}}.
\end{eqnarray}
Performing in Eq. (\ref{imaginary2}) angular integration we obtain
\begin{eqnarray}
I(R,\tau)=\frac{1}{4\pi}\int_0^{\infty} k\frac{e^{-E_k|\tau|}}{E_k}J_0(kR)dk
\end{eqnarray}
($J_0$ is the Bessel function of zero  order).
Using mathematical identity \cite{prudnikov}
\begin{eqnarray}
\int _0^{\infty}x\frac{e^{-p\sqrt{x^2+z^2}}}{\sqrt{x^2+z^2}}J_{0}(cx)dx=\frac{e^{-z\sqrt{p^2+c^2}}}{(p^2+c^2)^{1/2}},
\end{eqnarray}
we obtain
\begin{eqnarray}
I(R;\tau)=\frac{1}{4\pi v}\frac{e^{-\sqrt{v^2\tau^2+R^2}|m|/v}}{(v^2\tau^2+R^2)^{1/2}}.
\end{eqnarray}

In the gapless case $mR/v\ll 1$ we obtain \cite{cheianov}
\begin{eqnarray}
\label{gg}
{\cal G}(R;\tau)=-\frac{1}{4\pi}\frac{v\tau +i{\bf \gamma\cdot R} }{(v^2\tau^2+R^2)^{3/2}}.
\end{eqnarray}
Taking into account that $\{\gamma^i,\gamma^j\}=-2\delta_{ij}$ ($i=1,2$), and performing  integration in Eq. (\ref{abr})
 we obtain \cite{saremi}
\begin{eqnarray}
\label{final}
\chi^{CC}(R)=\frac{1}{64\pi vR^3},\qquad \chi^{AB}(R)=-\frac{3}{64\pi vR^3}
\end{eqnarray}
($CC$ can mean either $AA$ or $BB$).

In the opposite case $mR/v\gg 1$ we may approximate the GF as
\begin{eqnarray}
\label{gap25}
{\cal G}(R;\tau)=\frac{m\gamma^0R-i|m|{\bf \gamma\cdot R}}{4\pi v R^2} e^{-R|m|/v-v\tau^2|m|/2R}.
\end{eqnarray}
Performing  integration in Eq. (\ref{abr}),  both for intra-sublattice, and for inter-sublattice susceptibility we get
\begin{eqnarray}
\label{bu}
\chi=-\frac{1}{8v}\left(\frac{m}{\pi vR}\right)^{3/2}e^{-2R|m|/v}.
\end{eqnarray}

It is worth paying attention to the fact that Eq. (\ref{bu}) seems to contradict  rigorously proved theorem stating
that for any bipartite lattice at half filling, the RKKY interaction is antiferromagnetic
between impurities sitting on top of atoms belonging to opposite  sublattices (i.e., $A$ and $B$ sublattices in graphene), and is ferromagnetic between impurities
sitting on top of atoms
belonging to the same sublattice \cite{saremi,pereira,kogan}. However, the theorem is not applicable to Hamiltonian (\ref{dirac2}), with its last term meaning that if we rewrite  the Hamiltonian in the tight-binding representation, the
intra-sublattice hopping will appear, hence the lattice is no longer bipartite. More specifically, the spectrum still has the symmetry of that in bipartite lattice, but the wave functions do not \cite{kogan}.

\section{Imaginary frequency representation}

The approach will be based on equation \cite{abrikosov}
\begin{eqnarray}
\label{abr99}
\chi(R)=-\frac{1}{\pi}\int_{-\infty}^{\infty}{\cal G}^2(R;\omega)d\omega,
\end{eqnarray}
where ${\cal G}(R;\omega)$ is the GF in the coordinate -- imaginary frequency  representation.

\subsection{Gapped graphene}

The Green's function is
\begin{eqnarray}
\label{dirac26}
{\cal G}({\bf k},\omega)=\frac{-i\omega-v{\bf \gamma\cdot k}-\gamma^0m}{\omega^2+m^2+v^2k^2}.
\end{eqnarray}
Hence for the diagonal part of the GF we obtain
\begin{eqnarray}
\label{cc}
&&{\cal G}^{(d)}(R,\omega)=\frac{-i\omega-\gamma^0m}{2\pi}\int_0^\infty\frac{kJ_0(kR)dk}{\omega^2+m^2+v^2k^2}\nonumber\\
&&=\frac{-i\omega-\gamma^0m}{2\pi v^2}K_0\left(\sqrt{\omega^2+m^2} R/v\right),
\end{eqnarray}
and for the non-diagonal part
\begin{eqnarray}
\label{gap}
&&{\cal G}^{AB}(R,\omega)=\frac{-v}{2\pi}\int_0^\infty\frac{k^2J_1(kR)dk}{\omega^2+m^2+v^2k^2}\nonumber\\
&&=\frac{-\sqrt{\omega^2+m^2}}{2\pi v^2}K_1\left(\sqrt{\omega^2+m^2} R/v\right).
\end{eqnarray}
Substituting into Eq. (\ref{abr99}) we obtain \cite{dug}
\begin{eqnarray}
\label{gap3}
&&\chi^{CC}(R)=\frac{1}{4\pi^3 v^4}\nonumber\\
&&\int_{-\infty}^{\infty}(\omega^2-m^2)K_0^2\left(\sqrt{\omega^2+m^2} R/v\right)d\omega\nonumber\\
&&\chi^{AB}(R)=-\frac{1}{4\pi^3 v^4}\nonumber\\
&&\int_{-\infty}^{\infty}(\omega^2+m^2)K_1^2\left(\sqrt{\omega^2+m^2} R/v\right)d\omega.
\end{eqnarray}

For $m R/v\ll 1$, using   mathematical identity \cite{prudnikov}
\begin{eqnarray}
&&\int_0^{\infty}x^{\alpha-1}K_{\mu}(cx)K_{\nu}(cx)dx=\frac{2^{\alpha-3}}{c^{\alpha}\Gamma(\alpha)}\Gamma\left(\frac{\alpha+\mu+\nu}{2}\right)\nonumber\\
&&\Gamma\left(\frac{\alpha+\mu-\nu}{2}\right)\Gamma\left(\frac{\alpha-\mu+\nu}{2}\right)\Gamma\left(\frac{\alpha-\mu-\nu}{2}\right),
\end{eqnarray}
we recover Eq. (\ref{final}).

For $m R/v\gg 1$  we may use asymptotic expression for  modified Bessel functions
\begin{eqnarray}
K_{\nu}(z)\sim\sqrt{\frac{\pi}{2z}}e^{-z}.
\end{eqnarray}
After calculating the resulting integrals in Eq. (\ref{gap2}) using the Laplace method, we recover Eq. (\ref{bu}).

\subsection{Doped graphene}

For the case of doped (ungapped) graphene the GF is
\begin{eqnarray}
\label{matsu}
{\cal G}(k,\omega)=\frac{1}{i\omega+\mu-v{\bf \gamma\cdot k}},
\end{eqnarray}
where $\mu$ is a chemical potential.
From Eq. (\ref{matsu}) for the diagonal part of the GF we obtain
\begin{eqnarray}
\label{ty1}
&&{\cal G}^{CC}(R,\omega)=\frac{-i(\omega-i\mu)}{(2\pi)^2}\int_0^\infty\frac{kdk}{(\omega-i\mu)^2+v^2k^2} \nonumber\\
&&\cdot\int_0^{2\pi}e^{ikR\cos\theta}d\theta=\frac{-i(\omega-i\mu)}{2\pi}\int_0^\infty\frac{kJ_0(kR)dk}{(\omega-i\mu)^2+v^2k^2} \nonumber\\
&&=\frac{-i(\omega-i\mu)}{2\pi v^2}K_0\left[\text{sign}(\omega)(\omega-i\mu)R/v\right],
\end{eqnarray}
and for the non-diagonal part
\begin{eqnarray}
\label{ty2}
&&{\cal G}^{AB}(R,\omega)=\frac{-v}{(2\pi)^2}\int_0^\infty\frac{k^2dk}{(\omega-i\mu)^2+v^2k^2} \nonumber\\
&&\cdot\int_0^{2\pi}e^{ikR\cos\theta+i\theta}d\theta=\frac{-v}{2\pi}\int_0^\infty\frac{k^2J_1(kR)dk}{(\omega-i\mu)^2+v^2k^2} \nonumber\\
&&=\frac{-\text{sign}(\omega)(\omega-i\mu)}{2\pi v^2}K_1\left[\text{sign}(\omega)(\omega-i\mu)R/v\right].
\end{eqnarray}
($CC$ can mean either $AA$ or $BB$; $K_0$ and $K_1$ are the modified Bessel function of zero and first order respectively).
We have used mathematical identity, valid for Re $z>0$ \cite{prudnikov},
\begin{eqnarray}
\int_0^{\infty}\frac{x^{\nu+1}}{(x^2+z^2)^{\rho}}J_{\nu}(cx)dx=\frac{c^{\rho-1}z^{\nu-\rho+1}}{2^{\rho-1}\Gamma(\rho)}K_{\nu-\rho+1}(cz).\nonumber\\
\end{eqnarray}

The susceptibility (\ref{abr99})  is expressed through  the integrals
\begin{eqnarray}
\text{Re}\left\{\int_{0}^{\infty}(\omega-i\mu)^2K_{0,1}^2\left[(\omega-i\mu)R/v\right]d\omega\right\}.
\end{eqnarray}
Considering integrals in the complex plane it is convenient to deform the contour of integration and present the integrals as
\begin{eqnarray}
\text{Re}\left\{\int_{0}^{\infty}\omega^2K_{0,1}^2(\omega R/v)d\omega+\int^{0}_{-i\mu}\omega^2K_{0,1}^2(\omega R/v)d\omega\right\}.\nonumber\\
\end{eqnarray}
Taking into account the identity
\begin{eqnarray}
K_{\alpha}(-ix)=\frac{\pi}{2}i^{\alpha+1}[J_{\alpha}(x)+iY_{\alpha}(x)],
\end{eqnarray}
we  get \cite{sherafat}
\begin{eqnarray}
\label{mejer}
\chi_{\mu}^{CC}(R)=\chi_{\mu=0}^{CC}(R)\left[1-16\int_{0}^{k_FR}dzz^2J_{0}(z)Y_{0}(z)\right]\nonumber\\
\chi_{\mu}^{AB}(R)=\chi_{\mu=0}^{AB}(R)\left[1+\frac{16}{3}\int_{0}^{k_FR}dzz^2J_{1}(z)Y_{1}(z)\right],\nonumber\\
\end{eqnarray}
where $k_F=\mu/v$, and $\chi_{\mu=0}(R)$ are given by Eq. (\ref{final}).
The integrals in Eq. (\ref{mejer}) can be presented in terms of Meijer functions \cite{sherafat,gumbs} (I address the reader to
these References for the details).

It is interesting to compare the RKKY exchange in doped graphene, with its two sublattices and linear dispersion law,  with
that in ordinary two-dimensional electron gas. For the latter the Green's function is
\begin{eqnarray}
{\cal G}(k,\omega)=\frac{1}{i\omega+\mu-k^2/2m},
\end{eqnarray}
and the susceptibility turns out to be  \cite{fischer}
\begin{eqnarray}
\chi(R)\sim\frac{1}{R^2}\int_{0}^{k_FR}dzzJ_{0}(z)Y_{0}(z).
\end{eqnarray}

\section{Conclusions}

In the end we would like to mention again that in the case considered, the GF calculations in Euclidean space, as it is not infrequently happens,  have  advantages over those in Minkowskii space. In particular, using the former one have to operate only with the convergent integrals, in distinction to
what happens when one uses the latter.

\section{Acknowledgments}

Discussions with G. Gumbs, S. Kettemann, D. Loss,  M. Sherafati,  K. Szalowski, and K. Ziegler
were very illuminating for the author.

\section{Appendix: Approaches based on real frequency GF}

The  approach, used  in Refs. \cite{vozmediano,brey,saremi} is based on equation
\begin{eqnarray}
\label{abr9}
\chi(R)=\int \frac{d^2{\bf q}}{(2\pi)^2}\chi(\omega=0,{\bf q})e^{i{\bf q\cdot R}},
\end{eqnarray}
where
\begin{eqnarray}
\chi(\omega=0,{\bf q})=2\int\frac{d^2{\bf k}}{(2\pi)^2}\frac{n_F(\xi_{\bf k})-n_F(\xi_{{\bf k}+{\bf q}})}{E_{{\bf k}+{\bf q}}-E_{\bf k}};
\end{eqnarray}
$\xi_n=E_n-\mu$ and  $n_F(\xi)=\left(e^{\beta\xi}+1\right)^{-1}$ is the Fermi distribution function.
This approach, though looking quite straightforward, brings with it a problem. In  a model of infinite Dirac cones for
$\chi(\omega=0,{\bf q})$ we obtain a diverging integral. To obtain  finite values from these divergent integrals, as it was mentioned previously,  one has
to implement the complicated (and to some extent arbitrary) cut-off procedure \cite{saremi}.

The problem can be formulated in a different way. Being calculated in a  realistic band model, with the bands of finite width, $\chi(\omega=0,{\bf q})$
is not a universal quantity. It depends not only on infrared physics, but on the properties of electron spectrum and eigenfunctions in the whole Brillouin zone (even for small ${\bf q}$).

Another approach, formulated in Ref. \cite{dugaev}, starts from a well known equation for the susceptibility
\begin{eqnarray}
\label{dugaev}
\chi(R)= \frac{2i}{\pi}\int_{-\infty}^{\infty}G^2(R,E)dE,
\end{eqnarray}
where $G$ is the retarded green's function. Here again  the integral diverges on both limits of integration. However the authors changed the contour of integration, transforming the divergent integral (\ref{dugaev}) into the convergent integral  along the imaginary axis (see also Ref. \cite{duga}).   The authors also considered RKKY interaction in  gapped graphene, when the  power law  decrease of the  interaction with the distance turns into the exponential law. Actually,  the authors made the transition from real to imaginary frequencies, so some of our final results will be very close to ones obtained  in Ref. \cite{dugaev}.

The approach, using formula
\begin{eqnarray}
\chi(r,r')=\delta n(r)/\delta V(r')
\end{eqnarray}
and, hence, calculating electron susceptibility on the basis of equation
\begin{eqnarray}
\label{sherafati}
\chi(R)= -\frac{2}{\pi}\int_{-\infty}^{E_F}\text{Im}\left[G^2(R,E)\right]dE,
\end{eqnarray}
where  $E_F$ is the Fermi energy, was first used, in application to graphene to the best of our knowledge, in Ref. \cite{sherafati}.
An advantage of this approach is that it allows to easily consider the case of doped graphene, the disadvantage is that the approach, like  the one presented above, has to deal with the divergent integral (the integral with respect to $dE$ diverges at the lower limit of integration). Also in this case,
to obtain  finite values from these divergent integrals  one has
to implement the complicated (and to some extent arbitrary) cut-off procedure.


\begin{thebibliography}{99}


\bibitem{vozmediano} M. A. H. Vozmediano, M. P. Lopez-Sancho, T. Stauber and F. Guinea, Phys. Rev. B  {\bf 72}, 155121 (2005).

\bibitem{dugaev} V. K. Dugaev, V. I. Litvinov and J. Barnas, Phys. Rev. B  {\bf 74}, 224438 (2006).

\bibitem{brey} L. Brey, H. A. Fertig and S. D. Sarma, Phys. Rev. Let. {\bf 99}, 116802 (2007).

\bibitem{saremi} S. Saremi, Phys. Rev. B  {\bf 76}, 184430 (2007).

\bibitem{cheianov} V. V. Cheianov, O. Syljuasen, B. L. Altshuler, and V. Fal'ko, Phys. Rev. B {\bf 80}, 233409 (2009).

\bibitem{black} A. M. Black--Schaffer, Phys. Rev. B {\bf 81}, 205416 (2010).

\bibitem{sherafati} M. Sherafati and S. Satpathy, Phys. Rev. B  {\bf 83}, 165425 (2011).

\bibitem{uchoa} B. Uchoa, T. G. Rappoport, and A. H. Castro Neto, Phys. Rev. Lett. {\bf 106}, 016801 (2011).

\bibitem{bunder} J. E. Bunder and H.-H. Lin, Phys. Rev. B {\bf 80}, 153414 (2009).

\bibitem{power} S. R. Power and M. S. Ferreira,  Phys. Rev. B {\bf 83}, 155432 (2011).

\bibitem{kogan} E. Kogan, Phys. Rev. B {\bf 84}, 115119 (2011).

\bibitem{sherafat} M. Sherafati and S. Satpathy, Phys. Rev. B  {\bf 84}, 125416 (2011).

\bibitem{szalowski} K. Szalowski  Phys. Rev. B {\bf 84}, 205409 (2011).

\bibitem{bunder2} J. E. Bunder and James M. Hill, J. of Chem Phys {\bf 136}, 154504 (2012).

\bibitem{kettemann} H. Lee, J. Kim, E. R. Mucciolo, G. Bouzerar, S. Kettemann, Phys. Rev. B 85, 075420 (2012).

\bibitem{sherafati2} M. Sherafati and S. Satpathy, AIP Conf. Proc. 1461, pp. 24-33; doi:http://dx.doi.org/10.1063/1.4736868.

\bibitem{liwei} L. Jiang, X. Lu, W. Gao, G. Yu, Z. Liu, and
Y. Zheng, J. Phys.: Condens. Matter {\bf 24}, 206003 (2012).

\bibitem{loss} J. Klinovaja  and  D. Loss, 	arXiv:1211.3067v1.


\bibitem{gumbs} O. Roslyak, G. Gumbs, and D. Huang, arXiv:1211.5099v1.

\bibitem{katsnelson} M. I. Katsnelson, {\it Graphene: carbon in two dimensions}. Cambridge University Press (2012).

\bibitem{feynman} R. P. Feynman, {\it Quantum Electrodynamics} (Westview Press, 1961).

\bibitem{prudnikov} A. P. Prudnikov, Yu. A. Brychkov and O. I. Marichev, {\it Integrals and Series} Vol. 2  (Gordon and Breach Science Publishers, 1986).

\bibitem{pereira} V. M. Pereira, J. M. B. Lopes dos Santos, and A. H. Castro Neto,
Phys. Rev. B {\bf 77}, 115109 (2008).

\bibitem{abrikosov} A. A. Abrikosov, L. P. Gorkov, and I. E. Dzyloshinski, {\it Methods of Quantum Field Theory in Statistical Physics}, (Pergamon Press, 1965).

\bibitem{dug}  V. K. Dugaev, V. I. Litvinov and P. P. Petrov, Superlattices and Microstructures  {\bf 16}, 413 (1994).




\bibitem{fischer} B. Fischer and M. W. Klein, Phys. Rev. B {\bf 11}, 2025 (1975).


\bibitem{duga} V. I. Litvinov and V. K. Dugaev,  Phys. Rev. B  {\bf 58}, 3584 (1998).

\end{thebibliography}
\end{document}